\documentstyle[12pt]{article}
\begin{document}

\begin{flushright}
FERMILAB-Pub-95/059-A \\
December 1995
\end{flushright}

\vspace{1in}
\begin{center}
{\Large{\bf  On the formation of a Hawking-radiation photosphere around microscopic black holes}}\\

\vspace{.4in}

{\bf Andrew F. Heckler} \\
{\em  NASA/Fermilab Astrophysics Center,
Fermi National Accelerator Laboratory,}\\
{\em Batavia, IL 60510, USA}\\{\em email: aheckler@fnas04.fnal.gov}
 
\vspace{.2in}
\begin{abstract}
We show that once a black hole surpasses some critical temperature $T_{\rm crit}$, the emitted Hawking radiation interacts with itself and forms a nearly thermal photosphere. Using QED, we show that the dominant interactions are brems\-strahlung and electron-photon pair production, and we estimate $T_{\rm crit} \sim m_{e}/\alpha^{5/2}$, which when calculated more precisely is found to be  $T_{\rm crit} \approx $45 GeV. The formation of the photosphere is purely a particle physics effect, and not a general relativistic effect, since the the photosphere forms roughly $\alpha^{-4}$ Schwarzschild radii away from the black hole. The temperature $T$ of the photosphere decreases with distance from the black hole, and the outer surface is determined by the constraint $T\sim m_{e}$ (for the QED case), since this is the point at which electrons and positrons annihilate, and the remaining photons free stream to infinity. Observational consequences are discussed, and it is found that, although the QED photosphere will not affect the Page-Hawking limits on primordial black holes, which is most important for 100MeV black holes, the inclusion of QCD interactions may significantly effect this limit, since for QCD we estimate $T_{\rm crit}\sim \Lambda_{\rm QCD}$. The photosphere greatly reduces  possibility of observing individual black holes with temperatures greater than $T_{\rm crit}$, since the high energy particles emitted from the black hole are processed through the photosphere to a lower energy, where the gamma ray background is much higher. The temperature of the plasma in the photosphere can be extremely high, and this offers interesting possibilities for processes such as symmetry restoration.

\end{abstract}
\end{center}

\vspace{.6in}
submitted to {\em Physical Review D}
\renewcommand{\thesection}{\Roman{section}} 

\newpage

\section{Introduction}

As first shown by Hawking in 1975 \cite{Hawking75a},  quantum theory predicts that a black hole emits thermal radiation. The possibility of observing this thermal or "Hawking" radiation from, say, a solar mass black hole is impractically small: the entire black hole would emit only a few hundred quanta per second, and this is much too small of a flux to possibly be observed at astronomical distances.

However, since the temperature $T_{\rm BH}$ of the black hole, hence  flux of radiation, is inversely proportional to the black hole mass,  the possibility of detecting Hawking radiation from  much smaller mass black holes becomes observationally feasible. Page and Hawking, and several other authors \cite{Page76a,MacGibbon91a,Halzen91a} have placed upper limits on the density of very small mass (therefore very hot) black holes by constraining the total radiation produced to be less than the observed gamma ray background radiation. This method constrains the density of black holes with temperatures of order 100MeV. This particular number arises form the fact that a 100MeV black hole has a  lifetime on the order of the age of the universe, and although it is true that higher temperature black holes have higher fluxes, they also have much shorter lifetimes, and   the important quantity  (for background measurements) is  the time integrated flux. Therefore 100MeV black holes contribute the most to the gamma ray background.  We should note here that since there are no known astrophysical processes that can produce these small mass black holes, these constraints all assume they were produced in the early universe, hence they are called "primordial black holes".

An important issue in finding the constraints on the density of black holes is to determine the emission spectrum of the black hole. At first glance one might expect the observed spectrum of radiation from a black hole to be thermal, since the black hole emits thermal radiation from its surface (taking into account, of course, finite size effects). MacGibbon and Webber \cite{MacGibbon90a} have shown that the observed spectrum is not thermal simply because emitted particles such as quarks fragment into hadrons, photons, neutrinos etc., and this fragmentation plays a major role in determining the spectrum. 

There is another possibility, however, for affecting the observed spectrum. If the particles emitted from the surface {\em interact with each other} as they propagate away, then the spectrum observed far away from the black hole will not be the same as the emitted spectrum. In fact, if the particles interact strongly enough, then a photosphere will develop around the black hole, and the average energy of the particles at the outer surface of the photosphere will be much less than the average energy of particles emitted directly from the black hole. A similar effect also occurs in the sun, where the surface is much cooler than the central core, which produces the energy.

Previous authors have considered the possibility that the emitted particles do interact \cite{MacGibbon91a,Oliensis84a}, and they use (perhaps too) simplistic arguments that  the radiation emitted from the black hole interacts too weakly to form a photosphere. However,  using standard QED, we will show that brems\-strahlung (and photon-electron pair production) processes are important, and, for high enough black hole temperatures, the particles scatter and dramatically lose energy as they propagate away from the black hole. The principle idea is that at relativistic energies, the brems\-strahlung cross section is roughly constant (independent of energy). Since the density of emitted particles around the black hole increases with the black hole temperature, there will be a temperature at which brems\-strahlung (and pair production) scattering will become dominant.  Although this scattering is not enough to completely thermalize the emitted particles, a sort of near-thermal photosphere forms, and the average energy of the particles decreases dramatically.

The different sections in this paper are roughly self contained. In sections II and III, we discuss the subtleties of the brems\-strahlung cross section in some detail. In section IV, we apply this cross section and obtain a formula for the number of scattering ${\cal N}$ an average particle undergoes as it propagates to infinity, and find that for some critical black hole temperature, ${\cal N} =1$. Section V investigates the details of the photosphere, namely the radius of the inner and outer surface and the temperature and velocity of the fluid as as function of radius. At the end of this section and in section VI, the general characteristics of the photosphere are described. In section VII we discuss the the effect of including QCD interactions, and in section VIII we discuss the observational consequences of the photosphere.

In this paper we will neglect all general relativistic effects (except for Hawking radiation itself), because most of the interactions take place at a radius $r\gg r_{\rm BH}$, where $r_{\rm BH}$ is the radius of the black hole.

\section{Bremsstrahlung cross section}

For simplicity, let us first consider QED only.  In a later section we will consider the effects of QCD. 
In the relativistic limit, the differential cross section for brems\-strahlung in the center of momentum frame is found to be \cite{Jauch75a,Haug75a}
\begin{equation}\label{dsigma}
\frac{d\sigma(\omega)}{d\omega} \approx \frac{8 \alpha r_{0}^{2}}{E\omega }\left( \frac{4}{3}(E-\omega) + \frac{\omega^2}{E}\right)\left(\ln{\left[\frac{4E^2(E-\omega)}{m_{e}^{2}\omega}\right]} - \frac{1}{2}\right)
\end{equation}
where $\hbar = c = 1$, $r_{0} = \alpha/m_{e}$ is the classical electron radius, $E$ is the initial energy of an electron, and $\omega$ is the energy of the photon. The divergence of this differential cross section as $\omega \rightarrow 0$ is remedied by higher order corrections, which essentially impose an infrared cutoff.

In our application, we will not be concerned with this infrared divergence because low energy brems\-strahlung photons $\omega \ll E$, will not significantly contribute to the energy loss of the electrons. If we take an energy averaged cross section $\sigma = \int{\omega(d\sigma /d\omega)d\omega}/E$, we can obtain an approximate expression for the  cross section that is insensitive to the infrared divergence \cite{Jauch75a}
\begin{equation}\label{sigma}
\sigma_{\rm brem} \approx  8 \alpha r_{0}^2 \ln{\frac{2E}{m_{e}}}.
\end{equation}
This is in the center of mass frame. Examination of  the function $\omega(d\sigma /d\omega)$ reveals that the average energy lost in each collision is ${\ \lower-1.2pt\vbox{\hbox{\rlap{$<$}\lower5pt\vbox{\hbox{$\sim$}}}}\ } E$  \cite{Haug75a}.

The interesting and well known behavior of the the  relativistic brems\-strahlung (and pair production) cross section is that it does not decrease with energy. However, one must keep in mind that  the cross section is large compared to the energy scale of the particles, and there is a minimum interaction volume required in order for the process to take place. In order to understand this physically, consider that the brems\-strahlung process can be regarded as the following: two electrons  collide and exchange a photon, and one of the electrons is scattered into an off-shell state, and subsequently decays into an on-shell electron and a photon. The off-shell electron will travel a finite distance before it decays, therefore the process occurs in a finite volume. 

The amount of volume needed is simple to approximate: since the cross section is proportional to $m_{e}^{-2}$ and  is determined by the minimum momentum of the  off-shell electron.The off-shell momentum of the electron is, by conservation of energy and momentum, $p\cdot k$, where $p$ and $k$ are the four-momentum of the final electron and photon. Since $|p| \sim E$, $|k| \sim E$, and the average angle of photon emission is $\sim m_{e}/(2 E)$ \cite{Haug75a}, one obtains $p\cdot k\sim m_{e}^2$. Therefore the minimum scale size needed for this interaction to occur  is  $\sim m_{e}^{-1}$.We will use this result in the next section.

Note that along with the brems\-strahlung process ($e+e\rightarrow ee\gamma$), there is the similar process of photon-electron pair production ($e+\gamma \rightarrow ee^{+}e^{-}$). When we speak of brems\-strahlung, we will also tacitly include pair production because they have cross sections with the same functional dependence at relativistic energies \cite{Jauch75a} shown in eq.~(\ref{sigma}).

\section{Bremsstrahlung scattering : plasma effects}

In a pure QED theory, electrons, positrons and photons are emitted from a black hole that has a temperature $T_{\rm BH}>m_{0e}$, where $m_{0e}$ is the vacuum electron mass\footnote{The plasma mass, as defined below in this section, is always less than the vacuum mass at a radius $r>r_{\rm BH}$, when $T_{\rm BH} = m_{0e}$.}. We will assume that the energy spectrum of the particles emitted directly from the black hole is perfectly thermal, though in reality there are spin and finite source size effects \cite{MacGibbon90a}. We will make another simplifying approximation by noting that since the density of particles near the surface of the black hole is the thermal density at temperature $T_{\rm BH}$, then from conservation of particle number we obtain a formula for the density of particles at radius $r$:
\begin{equation} \label{n_{0}(r)}
n_{0}(r)\approx 2\int \frac{d^{3}p}{(2\pi)^3}f(p)\frac{r_{\rm BH}^{2}}{r^{2}}\approx \frac{1}{\pi^2}\frac{T_{\rm BH}}{(4 \pi)^{2}r^{2}}
\end{equation}
where the radius of the black hole $r_{\rm BH}=1/(4\pi T_{\rm BH})$, $f(p)=(\exp^{E(p)/T_{\rm BH}}\pm 1)^{-1}$ (for fermions and bosons), and recall that the $\theta$ angle should only be integrated to $\pi/2$, since the particles are being emitted from a surface,  the subscript $0$ is to remind us that we have not taken particle production from scattering into account. We will use this approximation for density for both photons and fermions. 

In calculating the density we must also take into account that brems\-strahlung and electron-photon pair creation are particle creating processes. Therefore with each scattering, more particles are being produced, and the particle density increases. Since the combinations $e^{\pm}+e^{\pm}\rightarrow e^{\pm}e^{\pm}\gamma$ and $e^{\pm}+\gamma \rightarrow ee^{+}e^{-}$ are all possible, let us simplify the picture and treat the electrons positrons and photons all as ``particles'' that undergo $2\rightarrow3$ body interactions: every two particles that interact create a third particle. Then, for ${\cal N}$ particle scatterings (see eq.~(\ref{Ndef})), we can account for the effect of particle creation by replacing the density the term $n_{0}$ by 
\begin{equation}\label{n(r)}
 n(r)= \left(\frac{3}{2}\right)^{{\cal N}(r)} n_{0}(r),
\end{equation}
which indicates that the density grows exponentially with each scattering.   

The relativistic  brems\-strahlung mean free path  of electrons and positrons at radius $r$ is
\begin{eqnarray}\label{lambda}
\lambda (r) &=& \frac{1}{n(r) \sigma_{\rm brem} |v|} \nonumber \\
& \approx&  \frac{m_{e}^2}{8 n(r) \alpha^3 ln{(2T_{\rm BH}/m_{e})}}
\end{eqnarray}
where we have approximated the  vector velocity difference between colliding particles to be $|v| \approx 1$, and the average energy of the electrons $E\approx T_{\rm BH}$.

Let us define ${\cal N}$ to be the number of scatterings  that an average particle has undergone as it travels from $r_{\rm min}$ to $r_{\rm max}$ from the black hole,
\begin{equation}\label{Ndef}
{\cal N} \equiv \int_{r_{\rm min}}^{r_{\rm max}}\frac{dr}{\lambda(r)}.
\end{equation}
Certainly if ${\cal N} > 1$, the particles  will begin to significantly interact, and possibly form a photosphere. 

In order to understand the important physics involved in this problem, let us first consider a naive calculation of ${\cal N}$, and assume that the electron mass is equal to the vacuum electron mass, i.e. $m_{e} = m_{0e}$, and that, since the minimum interaction volume is $m_{e}^{-3}$, we assume $r_{\rm min} = m_{e}^{-1}$ and $r_{\rm max} = \infty$. Using eqs.~(\ref{Ndef}) and (\ref{lambda}), and neglecting for the moment the density enhancement of eq.~(\ref{n(r)}), we obtain
\begin{eqnarray}\label{}
{\cal N}_{\rm brem} \approx \frac{ \alpha^{3}}{2\pi^4}\frac{T_{\rm BH}}{m_{0e}}\ln{\frac{2T_{\rm BH}}{m_{0e}}}.
\end{eqnarray}
In this naive picture, particle interaction becomes important (i.e. ${\cal N} {\ \lower-1.2pt\vbox{\hbox{\rlap{$>$}\lower5pt\vbox{\hbox{$\sim$}}}}\ } 1$) when
\begin{equation}\label{Tnaive}
T_{\rm BH} {\ \lower-1.2pt\vbox{\hbox{\rlap{$>$}\lower5pt\vbox{\hbox{$\sim$}}}}\ } 
\frac{\pi^2}{\alpha^3}m_{0e} \sim 20\,{\rm TeV},
\end{equation}
and most most of the particles scatter at a distance $m_{0e}^{-1}$ from the black hole.

However there are some problems with this naive picture. For example, at a distance $m_{0e}^{-1}$ from the black hole, the average inter-particle spacing at this temperature (\ref{Tnaive}) is $n^{-1/3} \approx \alpha m_{0e}^{-1}$, which is much smaller than the brems\-strahlung interaction length (it is even smaller than the electron compton wavelength), and one would expect this high density ``plasma'' to affect the cross section. A second and related problem is: what occurs in the region $r<m_{e}^{-1}$? Although the interaction distance of brems\-strahlung is $m_{e}^{-1}$, it does not mean that the interaction is turned off at shorter distances. It does mean, however that the cross section is suppressed. This is very similar to the LPM effect, where the brems\-strahlung cross section of a beam of electrons on a target is suppressed because the interaction distance is so large that the electrons can scatter via coulomb collisions during the brems\-strahlung interaction, and therefore suppress the brems\-strahlung cross section \cite{Landau53a}. 

Another, perhaps more lucid way of regarding the effect of brems\-strahlung suppression is to realize that the brems\-strahlung processes are not occurring in a vacuum, rather in a bath of radiation, or ``plasma''.  In our application we are calculating interaction rates which involve cross sections of particles in a  background plasma which is in a (almost) radially propagating collection of particles. When calculating quantities that include the masses of particles in a plasma, we must be careful to include the plasma mass of the particles. To a very good approximation, the plasma effects are easily included in cross sections by simply replacing the vacuum mass squared $m_{0}^{2}\rightarrow m_{0}^{2}+m_{\rm pm}^{2}$, where $m_{0}$ is the vacuum mass, and $m_{\rm pm}$ is the plasma mass. This can be easily understood by realizing that the finite temperature cross section is found by calculating feynman diagrams which are identical to the vacuum diagrams, but replacing every propagator by its finite temperature counterpart. The problem is then reduced to finding the finite temperature self energy, or in effect, finding the plasma mass. Let us define the total electron mass in a plasma to be
\begin{equation}\label{me}
m_{e}^{2}= m_{0e}^{2}+m_{\rm pm}^{2}(n,T,p),
\end{equation}
where we have explicitly noted that the plasma mass is a function of density, temperature and momentum.  The actual functional form of the electron plasma mass does not take this form, but it is accurate in the limit of very small or very large plasma mass (compared to the vacuum mass), and these are the limits that we will be interested in.

So the question is, what is the plasma mass of an electron propagating away from a black hole? The electron is propagating in a kind of plasma, but it is {\em not a thermal plasma}, since the density is not homogeneous, and the  particles are not moving in random directions, but rather they are propagating nearly radially  away from the black hole (recall the black hole is a small, but finite size source). The calculation of the plasma mass of a particle in an inhomogeneous, non-isotropic medium is non-trivial, and is beyond the scope of this paper. We will instead use a simple estimate for the plasma mass by noting that, for a thermal plasma with density $n$ and temperature $T$, $m_{\rm thermal}^{2}\approx 4\pi \kappa \alpha n/\bar{E}$ (see for example \cite{plasmamass}), where $\kappa$ is a constant of order unity, $\bar{E}$ is the average energy of the particles in the plasma, and we have made the rough approximation that $ 1/\bar{E}\approx \overline{1/E}$. We will therefore use the estimate that the plasma mass of a particle near a black hole is
\begin{equation}\label{mpm}
 m_{\rm pm}^{2} \approx  \frac{4\pi\alpha n(r)}{{\bar E}}\approx \frac{4\pi \alpha(3/2)^{2{\cal N}}}{a(4\pi^2 r)^{2}}
\end{equation}
where we have used eq.~(\ref{n(r)}) and we have assumed that since the average energy of a particle emitted from a black hole is $\sim a_{b}T_{\rm BH}$, where $a_{b}\approx 5$ \cite{MacGibbon90a}, and since the average energy of the a particle decreases with each scattering,  we estimate through conservation of energy that ${\bar E}\approx aT_{\rm BH}/ (3/2)^{\cal N}$.

\section{Onset of photosphere}

Let us consider a more accurate calculation brems\-strahlung scattering, which includes plasma effects discussed in the previous section. Combining eqs.~(\ref{n(r)}), (\ref{lambda}), and (\ref{Ndef}), we obtain a formula for number of  brems\-strahlung scatterings a particle undergoes travelling to a distance $R$ from the black hole:
\begin{equation}\label{Ndefbrem}
{\cal N}(R) \approx 8 \alpha^{3}\int_{r_{\rm BH}}^{R}{dr\left(\frac{ (3/2)^{{\cal N}(r)}n_{0}(r)}{m_{e}^{2}(r)} \ln{\left[\frac{2T_{\rm BH}}{m_{e}(r)}\right]}\right)},
\end{equation}
where we have explicitly shown the electron mass as a function of distance away from the black hole, since the plasma mass varies with density. 

Eq.~(\ref{Ndefbrem}) can be solved numerically (see Figure~1), but we can also determine solutions analytically  by approximating $(3/2)^{{\cal N}} \approx 1$, and approximating the argument of the logarithm to be $T_{\rm BH} /m_{0e}$ (because most of the integrand comes from the region where $m_{\rm pm} \sim m_{0e}$). Integrating from $r_{\rm BH}$ to infinity, we obtain
\begin{equation}\label{Nbrem}
{\cal N}_{\rm brem} \approx \left(\frac{a_{b}\alpha^{5}}{4\pi^3 }\right)^{1/2} \left(\ln{\frac{T_{\rm BH}}{m_{0e}}}\right) \frac{T_{\rm BH}}{m_{0e}} ,
\end{equation}
keeping in mind that this expression is valid only for ${\cal N}{\ \lower-1.2pt\vbox{\hbox{\rlap{$<$}\lower5pt\vbox{\hbox{$\sim$}}}}\ } 1$. Using the definition ${\cal N}(T_{\rm crit}) =1$, and approximating $a_{b}\approx 5$,  we obtain a critical temperature $T_{\rm crit}$ for the black hole
\begin{equation}\label{}
T_{\rm crit} \sim \left(\frac{(4\pi^{3}/a_{b})^{1/2}}{\ln{(\alpha^{-5/2})}}\right)\frac{m_{0e}}{\alpha^{5/2}} \sim 45 {\rm GeV}.
\end{equation}
Numerically solving (\ref{Ndefbrem}), we find that ${\cal N}_{\rm brem} =1$ when
\begin{equation}\label{Tmin}
T_{\rm crit} \simeq  45.2 {\rm GeV}.
\end{equation}
At this temperature, outgoing particles will scatter on average at least once via brems\-strahlung and photon-electron pair production processes. When the temperature increases above $T_{\rm crit}$, the particles scatter yet more, and at some point it is necessary to think of the plasma as no longer free streaming, but as an interacting fluid. 

We can also analytically estimate the behavior of ${\cal N}$ when it is much larger than unity by noticing that when one assumes ${\cal N}\gg 1$, one can use eqs.~(\ref{mpm},\ref{Ndefbrem}) to self consistently find  $m_{\rm pm} (r) \gg m_{0e}$ and 
\begin{equation}\label{Nlarge}
{\cal N}(r) \approx \frac{\ln{\left[8\alpha^{2} T_{\rm BH} r (\ln{3/2})\left(\ln{\frac{\pi^2}{\alpha^{5/2}}}\right)\right]}}{(\ln{3/2})},
\end{equation}
where we held ${\cal N}$ in the logarithm to be constant, then iteratively kept the highest term. Comparing to numerical results, this was found to be better than $1\%$ accurate for  ${\cal N} \gg1$.

\section{The thermal photosphere}

We have shown that when a black hole rises above a critical temperature $T_{\rm crit}$, the Hawking radiation particles  brems\-strahlung scatter many times as they propagate away from the black hole. As ${\cal N}$ gets large, the radiation will have to be described in a qualitatively different way. If the mean free path is short enough ($\lambda < r$), then a  kind of photosphere will form, and a fluid description is more appropriate. 

The unique environment  of  microscopic black hole photosphere, however, complicates the issue: the average energy and density of the particles is changing so fast as they propagate away from the black hole, that the particles never have enough time to fully thermalize. That is to say, the fluid in the photosphere is an imperfect fluid.

In order to show this, consider the following arguments which reveal an inconsistency with the perfect fluid description. Let us assume for the moment that the particles scatter enough that at some point they become a thermal, perfect fluid. In the rest frame of a fluid element, the time scale for particle scattering $\tau_{\rm mfp}$ must be smaller than the time scale for any change in macroscopic parameters (such as temperature) of the fluid element $\tau_{\rm fl}$. For a relativistic plasma at temperature $T$, $\tau_{\rm mfp} \sim (\alpha^{2} T)^{-1}$. In the rest frame of the black hole, the time scale for change in macroscopic parameters is$\sim r/v \sim r$ (this is also borne out by eq.~(\ref{Trgamma})). Consequently, boosting to the frame of the fluid, we find $\tau_{\rm fl} \approx r/\gamma)$.  Therefore, the perfect fluid condition $\tau_{\rm mfp}< \tau_{\rm fl}$, becomes
\begin{equation}\label{fluid}
\frac{\alpha^{2}Tr}{\gamma} > 1.
\end{equation}
Since we have assumed that the fluid is a perfect gas, we can determine other relations between $T$, $\gamma$, and $r$ by simply employing conservation of energy-momentum ($\partial_{\mu}T^{\mu\nu}=0$) and entropy ($\partial_{\mu}(su^{\mu})=0$) using spherical symmetry, and assuming that the fluid obeys the relativistic equation of state ($p=\rho/3)$ :
\begin{equation}\label{Trgamma}
\frac{T_{1}}{T_{2}}=\frac{\gamma_{2}}{\gamma_{1}}=\frac{r_{2}}{r_{1}}\frac{u_{2}^{1/2}}{u_{1}^{1/2}},
\end{equation}
where $\gamma=(1-u^{2})^{-1/2}$ is the Lorentz gamma factor of the fluid, and $T_{1}$ is the temperature of the fluid at radius $r_{1}$, etc.

The inconsistency comes from the following observation: if the particles interact just enough to meet the perfect fluid constraint (eq.~(\ref{fluid})) at some radius $r_{0}$, then according to eq.~(\ref{Trgamma}), the constraint is subsequently violated at any $r > r_{0}$, and the fluid is no longer considered to be perfect. However, if one were to look at the microscopic interactions, occurring at $r$, one would find copious scattering (increasing with $r$), which leads one back to assuming a perfect fluid. The solution to this inconsistency lies in realizing that the fluid is not perfect and eq.~(\ref{Trgamma}) is incorrect. Another way of stating this is that the temperature and velocity  of the fluid change significantly over a mean free path, so there is a large amount of entropy production ($\partial_{\mu}(su^{\mu})> 0$).

In order to find a solution to this problem, which ideally would be solved via the Boltzmann equation, we will make a simplification based on the inconsistency stated above. That is, since the perfect fluid equations (\ref{Trgamma}) will not allow for $\alpha^2 rT/\gamma \gg1$, and the microscopic equations (\ref{Nlarge}) will not allow for $\alpha^2 rT/\gamma \ll1$, we will then assume
\begin{equation}\label{balance}
\frac{\alpha^{2}Tr}{\gamma} = 1.
\end{equation}
We will employ this constraint along with conservation of energy in order to find $T(r)$ and $\gamma(r)$, but first let us discuss the general form of the photosphere.  One should keep in mind that even though we can identify $T$ as approximately the average energy of of the particles in the rest frame of the fluid, we use the term ``temperature'' loosely since the fluid is not a perfect fluid.

The general form of the photosphere will be a thick shell of plasma (see Fig.~2). The fluid at the  inner surface of the photosphere shell will have some  ``temperature'' $T_{0}$, and as a fluid element of the photosphere propagates outward, it will cool until it eventually reaches a temperature below the electron mass. This is the outer surface of the photosphere: at this point the electrons and positrons will annihilate, and the photons will free stream away to infinity. This picture is similar to the fireball model  of Goodman \cite{Goodman86a}. 

The radius $r_{0}$, temperature $T_{0}$, and velocity $u_{0}$ of  the fluid of the inner surface of the photosphere are found from eq.~(\ref{balance}) and the conservation of energy-momentum at the boundary of the inner surface of the photosphere. We will make the approximation that on the inside of the boundary, there is  free streaming radiation, and on the outside of the boundary there is a near-thermal plasma at temperature $T_{0}$, and we will assume the boundary is a thin layer. Thus we will apply the conservation of energy-momentum $\partial_{\mu}T^{\mu\nu}=0$ for a planar surface. The energy momentum tensor for radiation free steaming (in one direction) is 
\begin{equation}\label{Tfs}
T_{\rm fs}^{\mu\nu} = \rho_{\rm fs}\frac{k^{\mu}k^{\nu}}{k_{0}^{2}},
\end{equation}
where $\rho_{\rm fs}$ is the energy density of the free steaming radiation at radius $r_{0}$ from the black hole, and $k^{\mu}$ is the (averaged) momentum four-vector of a particle of radiation. The fluid in the photosphere is an imperfect gas, but let us assume that the boundary conditions will not be much affected if we assume for the moment that the fluid is perfect. For a perfect gas, $T_{\rm per}^{\mu\nu} = (\rho_{0} +p_{0})u_{0}^{\mu}u_{0}^{\nu} - p_{0} g^{\mu\nu}$. By assuming a steady state, one obtains two equations 
\begin{eqnarray}\label{twoeqs}
\rho_{\rm fs}&=&(\rho_{0}+p_{0})\gamma_{0}^{2}u_{0} \nonumber \\
\rho_{\rm fs}&=&(\rho_{0}+p_{0})\gamma_{0}^{2}u_{0}^{2} + p_{0},
\end{eqnarray}
for which one can solve for the velocity of the perfect gas at the inner boundary of the photosphere:
\begin{equation}\label{v1}
u_{0} = 1/3,
\end{equation}
and the formula for the energy density of the fluid, in the rest frame of the fluid 
\begin{equation}\label{rho1}
\rho_{0}= 2\rho_{\rm fs}=\frac{2aT_{\rm BH}^{2}}{(4\pi^2)^{2}r^{2}},
\end{equation}
where, as stated earlier $a$ is defined as $a= \rho/(n T)\approx 5$, and we have once again assumed a relativistic equation of state. Using the fact that $\rho_{0} = a T_{0} n \approx (2a/\pi^2) T_{0}^4$, and using eqs~(\ref{v1}) and (\ref{fluid}), we obtain an estimate of  the parameters of the inner surface of the photosphere
\begin{equation}\label{inner}
r_{0}=\frac{4\pi}{\gamma_{0}^{2}\alpha^{4} T_{\rm BH}}, \,\,\,\,\,\,\,\,\,\,\,T_{0}= \frac{\gamma_{0}\alpha^{2}}{4\pi}T_{\rm BH},\,\,\,\,\,\,\,\,\,\,\,{\rm and} \,\,\,\,\,\,\,\,\,\,\, \gamma_{0}=\frac{9}{8}.
\end{equation}
Notice that the size and temperature of the photosphere are functions of the black hole temperature $T_{\rm BH}$. 

We can then determine $T$ and $\gamma$ at larger radii by simply using eq.~(\ref{balance}) and the conservation of energy. Assuming $u\approx 1$, the conservation of energy requirement in the photosphere becomes roughly
\begin{eqnarray}\label{consE}
r^2 \gamma^2 T^4 \approx r_{0}^{2} \gamma_{0}^{2} T_{0}^{4},
\end{eqnarray}
and with eq.~(\ref{inner}) we find
\begin{eqnarray}\label{imperfect}
T(r)= \left( \frac{\gamma_{0}T_{\rm BH}}{4\pi \alpha^2 r^2} \right)^{1/3}\,,\,\,\,\,\,\,\,\, \gamma(r)= \left( \frac{\gamma_{0}\alpha^4 T_{\rm BH} r}{4\pi} \right)^{1/3}.
\end{eqnarray}

The temperature of the photosphere decreases as the radius increases, and eventually the temperature will decrease to $m_{0e}$. This is the outer boundary of the photosphere since most of the electrons and positrons will annihilate at this point, and the mean free path will become so large that the particles will simply free stream to infinity. Using eqs.~(\ref{Trgamma}) and (\ref{inner}) we find the parameters of the  outer  surface of the photosphere to be
\begin{eqnarray}\label{outer}
r_{p}&=& \left( \frac{\gamma_{0}T_{\rm BH}}{4\pi m_{0e}} \right)^{1/2} \frac{1}{\alpha m_{0e}}, \,\,\,\,\,\,\,\,\,\,\,T_{p} = m_{0e},\,\,\,\,\,\,\,\,\,\,\, \nonumber \\
& &{\rm and} \,\,\,\,\,\,\,\,\,\,\, \gamma_{p}= \alpha \left( \frac{\gamma_{0}T_{\rm BH}}{4\pi m_{0e}} \right)^{1/2}.
\end{eqnarray}
In terms of critical temperature $T_{\rm crit}$,
\begin{eqnarray}\label{}
r_{p} \sim \frac{1}{\alpha^2 m_{0e}}\left( \frac{T_{\rm BH}}{T_{\rm crit}} \right)^{1/2} \,,\,\,\,\,\,\,\,\,\,\,\,\,\, \gamma_{p} \sim \left( \frac{T_{\rm BH}}{T_{\rm crit}} \right)^{1/2}
\end{eqnarray}
The average energy observed far away from the black hole (beyond the photosphere) is roughly
\begin{equation}\label{Eobs}
\bar{E}_{\rm obs} \approx a_{th} \gamma_{p} T_{p} \sim m_{0e}\left( \frac{T_{\rm BH}}{T_{\rm crit}} \right)^{1/2}, 
\end{equation}
where $a_{\rm th} \approx 3$ is obtained from a pure thermal plasma, as opposed to $a_{b}\approx 5$, which is obtained from the black hole spectrum. Figure~2 shows the average energy an observer beyond the photosphere will see for a given temperature of the black hole inside. 

Since the density of photons at $r_{p}$ in the rest frame of the black hole is $\sim \gamma_{p}m_{0e}^3$,  we can estimate the total number of photons emitted by the black hole per second
\begin{equation}\label{Ndot}
\dot{N}_{\rm photons} \sim \frac{ T_{\rm BH}}{\alpha^2} \left( \frac{T_{\rm BH}}{T_{\rm crit}} \right)^{1/2},
\end{equation}
which is roughly $(T_{\rm BH}/T_{\rm crit})^{1/2}/\alpha^{2}$ times greater than the particle emission rate directly from the black hole. See figure~2.

One can also estimate the number of electrons and positrons left over after interactions between them have ``frozen out'' from the rapid decrease in the photosphere temperature. The freeze out density is determined by comparing the mean free path to the radius, as before. If we approximate the $e^{\pm}$ density in the rest frame of the fluid as $n_{e^{\pm}}'\sim T^3 e^{-m_{0e}/T}$, then the mean free path in the rest frame of the black hole is $\lambda \sim e^{m_{0e}/T}/(\gamma \alpha^2 T)$. If we impose $\lambda \geq r$ at the outer surface of the photosphere, we find $n_{e^{\pm}}' \sim m_{0e}^3(T_{\rm crit}/T_{\rm BH})^{1/2}/\gamma_{p}$. Therefore, in the rest frame of the black hole we find  at the outer surface of the photosphere that the freezing-out density of $e^{\pm}$ is $n_{e^{\pm}} \sim m_{0e}^3(T_{\rm crit}/T_{\rm BH})^{1/2}$, and the total number of $e^{\pm}$ emitted per second by the photosphere is
\begin{equation}\label{}
\dot{N}_{e^{\pm}} \approx \frac{(4\pi) m_{0e}}{\alpha^4}\left( \frac{T_{\rm BH}}{T_{\rm crit}} \right)^{1/2},
\end{equation}
which is roughly a factor of about $T_{\rm BH}/T_{\rm crit}$ less than the photon emission rate.

\section{The big   picture}

Now we can describe the whole picture of the photosphere which forms around a  hot black  hole (see Figure~3). At black hole temperatures $T_{\rm BH}<T_{\rm crit} \sim $ 30GeV, there is no scattering and no photosphere-- the particles free stream away from the black hole, and the far away observer sees the thermal spectrum directly emitted by the black hole (we are neglecting the fragmentation of quarks etc \cite{MacGibbon90a}). 

As the black hole temperature rises to $T_{\rm BH} {\ \lower-1.2pt\vbox{\hbox{\rlap{$>$}\lower5pt\vbox{\hbox{$\sim$}}}}\ }T_{\rm crit}$, brems\-strahlung and photon-electron pair production begins to occur. The particles will begin to scatter (${\cal N} \sim 1$) via brems\-strahlung and pair production at a radius of $r=(8 \alpha^{2}T_{\rm BH})^{-1}$, and as they continue to scatter, they will eventually form a thermalized photosphere at a radius $r _{0} \sim 4\pi /(\alpha^4 T_{\rm BH})$. The characteristics of the photosphere are described in the previous section, and represented in Figure~2 and Figure ~3.

\begin{table}
\begin{center}
\begin{tabular}{|c||c|c|c|c|c|}		
\hline
&$T_{\rm BH}$ & $\bar{E}_{\rm obs}$ & $r_{\rm ph}$ & $\tau_{\rm BH}$ & $M_{\rm BH}$\\  \hline 
{\lower 2pt \hbox{Photosphere}} &  {\lower 8pt \hbox{45GeV}} &  {\lower 8pt \hbox{$m_{0e}$}}& {\lower 8pt \hbox{$(\alpha^2m_{0e})^{-1}$}} & {\lower 8pt \hbox{$10^8$ sec}} &{\lower 8pt \hbox{$10^{11}$ gr}} \\ 
{\raise 2pt \hbox{initially forms}} & & & & &  \\ \hline
		& & & & &  \\
$T_{\rm BH} =$10 TeV & $10$ TeV & $10$ MeV  & $10^{-5}$ cm & $10$ sec & $10^{9}$ gr \\
		& & & & & \\ \hline
		& & & & &  \\
$r_{\rm ph}=c\tau_{\rm BH}$ & $10^{9}$ GeV & $1$ GeV  & $10^{-3}$ cm & $10^{-13}$ sec & $10^{4}$ gr \\
		& & & & & \\ \hline
{\lower 2pt \hbox{Photosphere}}  & {\lower 8pt \hbox{--}} & {\lower 8pt \hbox{$10^{8}$ GeV}} & {\lower 8pt \hbox{$10^2$cm}}  & {\lower 8pt \hbox{--}}& {\lower 8pt \hbox{--}} \\
{\raise 2pt \hbox{finally dissipates}} & & & & &  \\   \hline
\end{tabular}
\caption{Characteristics of black hole and QED photosphere at various times. Note that the average particle energy $\bar{E}_{\rm obs}$ observed far from the black hole (beyond the photosphere)  is much smaller than the black hole temperature. Note also that when the black becomes so hot that  $r_{\rm ph}=c\tau_{\rm BH}$, the black hole quickly evaporates and leaves behind a remnant photosphere, which eventually dissipates. All values are approximate.}
\end{center}
\end{table}

As the black hole continues to increase in temperature, the  photosphere grows. For low enough black hole temperatures, we can treat the photosphere as steady state--that is the effect of the changing black hole temperature is negligible. Once the black hole temperature increases to the point where its lifetime $\tau_{\rm BH}$ is equal to the light crossing time of the photosphere, that is when
\begin{equation}\label{}
r_{\rm ph}=c\tau_{\rm BH},
\end{equation}
then the photosphere is no longer in a steady state. Beyond this point the black hole will evaporate quicker than the photosphere dissociates, and  there will be a period when the there is a remnant photosphere, with no black hole in the center. 

The lifetime of the black hole is found from its rate of mass loss: 
\begin{equation}\label{Mdot}
\dot{M}_{\rm BH}\approx  -4\pi r_{\rm BH}^{2}T_{\rm BH}^{4}.
\end{equation}
Strictly speaking, one must also include a factor of the relevant degrees of freedom such as quarks, gluons etc.\cite{MacGibbon90a}, but since we are only including the presence of photons, electrons and positrons, we will neglect this factor. The lifetime $\tau_{\rm bh}$ of the black hole is then
\begin{equation}\label{taubh}
\tau_{\rm BH}\approx \frac{M_{\rm pl}^{2}}{6T_{\rm BH}^3 },
\end{equation}
where $M_{\rm pl}$ is the Planck mass. 

The important features of the black hole and its photosphere are summarized in Table~1 for various times during its evolution. For the final moments of the black hole, we are assuming a naive picture that the black hole simply evaporates into radiation, with a maximum temperature of $M_{\rm pl}$.

\section{Including QCD}

The brems\-strahlung process also occurs in QCD. In this case, two quarks collide and emit a gluon, and the coupling $\alpha_{S}$ is much larger than the QED coupling. Adding quarks, gluons and QCD interactions to the picture will therefore change the critical temperature, size, and structure of the photosphere. 

In order to show this, let us use a simple model of QCD. First of all, because of  the asymptotic freedom of QCD interactions, the average interparticle spacing must be less than $\Lambda_{\rm QCD}$,
\begin{equation}\label{qcdlimit}
n^{-1/3}(r) < \Lambda_{\rm QCD},
\end{equation}
in order  for the coupling constant to be small enough that perturbation theory is valid. When the interparticle spacing is larger than this, the quarks and gluons simply form hadrons etc, which is described by MacGibbon and Webber \cite{MacGibbon90a}.

For simplicity, let us assume that the  QCD brems\-strahlung cross section has the exact same form as the QED cross section, only one must now use the strong coupling $\alpha_{S}$, and the mass of the quark involved $m_{q}$,
\begin{equation}\label{sigmaqcd}
\sigma^{\rm QCD}_{\rm brem} \sim  \frac{8 \alpha^{3}_{S}}{m_{q}^2} \ln{\frac{2E}{m_{q}}}.
\end{equation}
Once again, we must take into account the plasma mass of the quark, and we do so by following the same prescription as for the plasma mass of the electron (\ref{mpm}). We also will simplify the problem by estimating that the smallest mass of the quark is $\Lambda_{\rm QCD}$. Therefore we estimate 
\begin{equation}\label{mq}
m_{q}^2 = \Lambda_{\rm QCD}^2 + \frac{\alpha_{S}({\cal N}^{\rm QCD} +1)^2}{(4\pi^2r)^2}.
\end{equation}
This estimate will always insure that the important scale is the interparticle spacing, and that the smallest energy scale is $\Lambda_{\rm QCD}$. Following the same procedure as in the QED case, we estimate that the critical temperature at which quarks begin to QCD brems\-strahlung scatter ({\em i.e.} ${\cal N}^{\rm QCD} =1$)
\begin{equation}\label{}
T_{\rm crit}^{\rm QCD} \sim \frac{\Lambda_{\rm QCD}}{\alpha_{S}^{5/2}} {\ \lower-1.2pt\vbox{\hbox{\rlap{$>$}\lower5pt\vbox{\hbox{$\sim$}}}}\ } \Lambda_{\rm QCD}.
\end{equation}
The true value of $T_{\rm crit}^{\rm QCD}$ is difficult to estimate, since $\alpha_{S}$ is such a sensitive function of energy at these scales. Clearly  the scale $\Lambda_{\rm QCD}$ will be important, but a detailed model of QCD is needed to obtain a more accurate number. Since $T_{\rm crit}^{\rm QED} > T_{\rm crit}^{\rm QCD} \sim 100$MeV , it is also clear that QCD will play an important in the formation of a photosphere around a microscopic black hole.

\section{Consequences for Observation}

As stated in the introduction, the Page-Hawking limit, which constrains the total flux of Hawking radiation  from all of the evaporating black holes to be less than the observed gamma ray background \cite{Page76a,MacGibbon91a,Halzen91a},  most stringently constrains black holes emitting  100MeV photons. Since  $T_{\rm crit}^{\rm QED} \sim 45$GeV, the QED photosphere will not have a large effect on this constraint. Even as the temperature of the black hole increases to the point where the photosphere begins to emit 100MeV photons again (when $T_{\rm BH} \sim (10^4)T_{\rm crit}$, see figure~2), the black hole will have such a small mass compared to its mass when it was a100MeV black hole, that its contribution to the background (when it is converted to photons) at this point will be negligible. Therefore the QED photosphere will have a negligible affect on Page-Hawking limit.

{\em However}, since $T_{\rm crit}^{\rm QCD} \sim \Lambda_{\rm QCD} \sim 100$MeV, the QCD photosphere may  play a dominant role in determining the number 100MeV photons emitted by a black hole. Clearly, a full calculation including QCD interactions is needed in order to determine the energy spectrum of particles coming from the photosphere. In general, one would expect an added flux of particles starting at energies $\sim \Lambda_{\rm QCD}$. Furthermore, compared to the previous calculations of the spectrum \cite{MacGibbon90a}, there will be much smaller fluxes at energies $E>T_{\rm crit}$ because in the standard calculation, one expects a black hole with $T_{\rm BH} \sim E$ to produce particle of energy $E$, whereas, due to the photosphere, these fluxes are actually produced by black holes with $T_{\rm BH} \sim  E^2/(m_{0e}\alpha^{5/2})$,  which have masses $\sim E/(m_{0e}\alpha^{5/2})$ smaller  than expected from the standard calculation, and so the total energy that these smaller mass black holes can contribute to the gamma ray background at energy $E$ is $E/(m_{0e}\alpha^{5/2})$ times less than the standard calculation.

Another way of constraining the density black holes is based on the fact that we have not, as far as we know, observed an individual black hole evaporating in its final stages \cite{MacGibbon91a,Halzen91a,Alexandreas93a}. For example, Halzen {\em et. al.} \cite{Halzen91a} show, using the thermal radiation plus quark fragmentation model, that  if a a black hole with $T_{\rm BH} = 100$GeV, which has a lifetime of about $10^{7}$ seconds and radiates 100GeV photons is closer than about 1 parsec, then its emission will be above the 100 GeV background, and can be observed. 

If we apply the effect of the photosphere to the standard constraints on the distance individual black holes, we find that, since the QED photosphere decreases the energy of the particles emitted from the black hole, observation becomes much more difficult because the background is much higher at lower energies. To illustrate this, let us use the example of the $T_{\rm BH} =100$GeV black hole. The observer at infinity will only see photons that have been processed through the photosphere, and as a result, will only see photons with an average energy of ${\bar E}_{\rm obs}\sim m_{0e}(T_{\rm BH}/T_{\rm crit})^{1/2}\sim 1$MeV. In order to conserve energy, the photon flux will consequently increase by a factor of $T_{\rm BH}/{\bar E}_{\rm obs}$ over the standard no-photosphere assumption. But the the observed gamma ray background flux  is proportional to $E^{-2.5}$ \cite{MacGibbon91a}. In order to see the 100GeV black hole, one must look in the 1MeV energy range, where the background is much higher. Even though the black hole flux is increased by a factor  of  $T_{\rm BH}/{\bar E}_{\rm obs}$ by the photosphere, this is not nearly enough to compensate for the increase in background. Therefore, from this point of view, the photosphere makes the observation of individual black holes much more difficult, and the present limits must be reconsidered. That is to say, individual primordial black holes may be a lot closer than the present constraints prescribe. 

The presence of the photosphere will change the constraints on individual black holes in several other important ways. If we consider observing the sky at some energy $E_{\rm obs}$, then for some range of energies $m_{0e} {\ \lower-1.2pt\vbox{\hbox{\rlap{$<$}\lower5pt\vbox{\hbox{$\sim$}}}}\ } E_{\rm obs} {\ \lower-1.2pt\vbox{\hbox{\rlap{$<$}\lower5pt\vbox{\hbox{$\sim$}}}}\ } T_{\rm crit}$ there will be black holes at {\em  two different} temperatures which will both produce photons of average energy $E_{\rm obs}$ and contribute to a signal.  For example, if $E_{\rm obs} =$10MEV, then a black hole with $T_{\rm BH} \sim$10MeV, which has no  photosphere, and a black hole with $T_{\rm BH} \sim$10TeV, which has a (QED) photosphere will both radiate photons $\bar{E}_{\rm obs}\sim$10MeV (see Fig.~2). The 10TeV black hole, however, will emit photons at a {\em much} higher rate than the 10MeV black hole. In particular, in the no-photosphere case, $E_{\rm obs} \sim T_{\rm BH}$ and the total photon flux is  of order $E_{\rm obs}$; however with the photosphere, we find from eqs.~(\ref{Eobs},\ref{Ndot}) that the photon flux is roughly 
\begin{equation}\label{}
\dot{N} \sim \left( \frac{E_{\rm obs}}{T_{\rm crit}} \right)^{3}\frac{T_{\rm crit}}{\alpha^{19/2}}
\end{equation}
which is much greater than the flux from the lower temperature black hole.

However, one should note that the 10TeV black hole has a lifetime of only about 10 seconds, which severely limits the integration time of the observation. This exemplifies an important observational consequence of the photosphere. The photosphere decreases the average energy of the particles to such an extent that very high energy photons ($\bar{E}_{\rm obs}$) can only be produced by extremely high temperature black holes, which have such extremely short lifetimes, that they are, practically speaking, unobservable (see Table~1). This will dramatically weaken the constraints  made by high energy observations such as in Ref.~\cite{Alexandreas93a}. 

Because the photosphere decreases the energy of the emitted particles, the possibility that black holes are the source of ultra-high energy background photons (or other cosmic rays) seems remote. Even by including QCD and electro-weak theory, it would be difficult to produce ultra high energy $e^{\pm}$ or photons that would not be processed in the photosphere. Of course, the black hole could emit other high energy particles such as neutrinos, but even in this case a  neutrino photosphere will eventually form, for hot enough black hole temperatures.

\section{Conclusions and speculations}

The main purpose of this paper is to show that by using simple QED theory, one can show that a photosphere does indeed form around a black hole, and that this can have important observational effects. In  order to understand the full consequences of the photosphere, however, one must include QCD interactions, which will cause the photosphere to form at lower black hole temperatures. 

Nonetheless, even with the inclusion of QCD one might expect from our analysis that the Page-Hawking limit on the density of primordial black holes  will at most be affected by only an order of magnitude or two.  The observation of individual black holes, on the other hand, will be dramatically affected by the photosphere. The photosphere makes observing individual black holes much more difficult, and this opens up the possibility that an individual black hole can be much closer than previous constraints prescribe.

Another interesting aspect of the black hole photosphere comes from the fact that it is a unique astrophysical environment. The plasma in the photosphere can reach extremely high temperatures (recall $T_{0} \sim \alpha^2 T_{\rm BH}$), and one would expect interesting physical processes to occur in such an environment. For example, at very high energies, one would expect symmetry restoration to occur, and this opens up a wide range of possibilities. One simple example is baryogenesis. Since the photosphere can reach temperatures well above the electroweak scale, one would expect baryon violating interactions to occur via the electro-weak anomaly. Furthermore, since the photosphere is naturally out of thermal equilibrium, all of the requirements for baryogenesis are present \cite{Sakharov67a}, and the black hole should produce a net baryon number in its photosphere. One would also expect phase transitions such as the QCD phase transition to occur in the photosphere. Clearly, the inclusion of other theories will make the black hole photosphere much more complicated than the simple QED photosphere.

I would like to thank Eric Braaten for discussions about the plasma mass, and I would also like to thank Craig Hogan, Scott Dodelson and Chris Hill for many helpful discussions. This work was supported in part by the DOE and by NASA (NAG5-2788) at Fermilab, and NAG5-2793 at the University of Washington.

\listoffigures

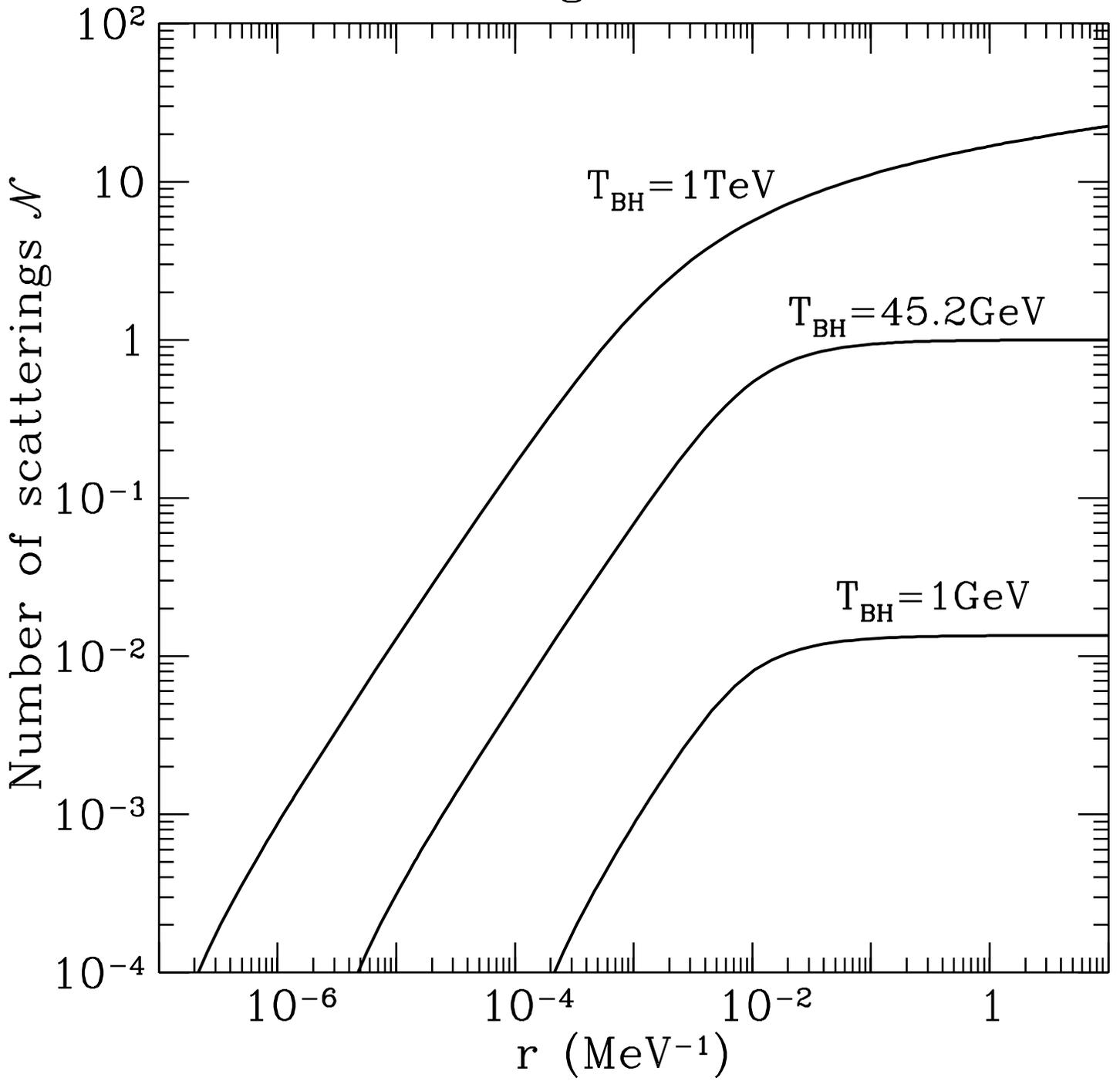
Figure 1. The average number of brems\-strahlung or photon-electron pair production scatterings ${\cal N}$  that an $e^{\pm}$ or photon undergoes as it propagates a distance $r$ away from the black hole, for various black hole temperatures $T_{\rm BH}$. Note that ${\cal N}$ approaches unity for $T_{\rm BH} = T_{\rm crit} \simeq $45.2GeV.\\

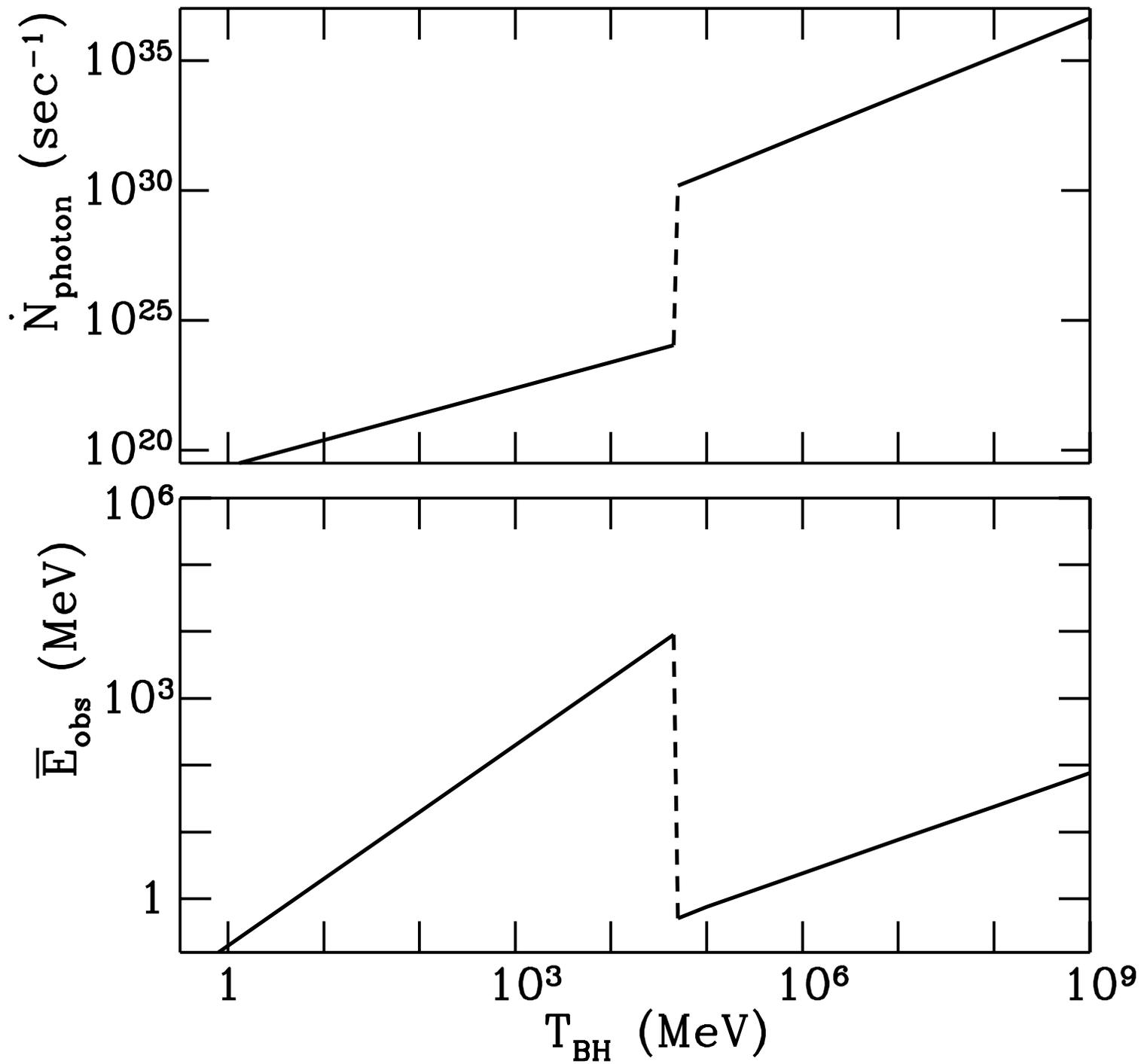
Figure 2. The total photon flux and average photon energy that a far-away observer will see as a function of the black hole temperature $T_{\rm BH}$. The dashed line is an extrapolation of what occurs at the critical temperature $T_{\rm crit}^{\rm QED} \approx 45$GeV. The (local) maximum in the observed energy is at about $T_{\rm crit}$, and the minimum at about $m_{0e}$. 
\\

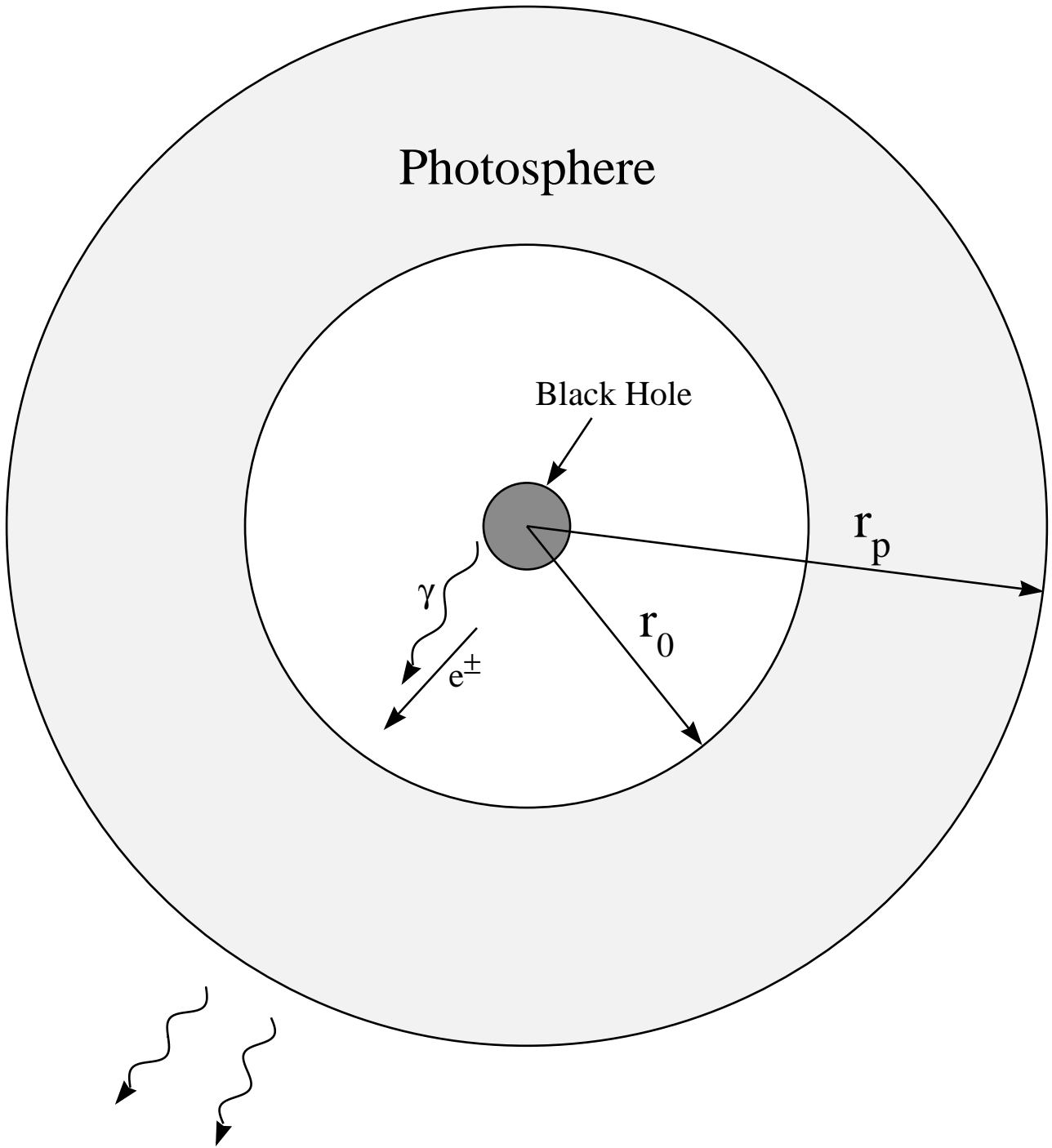
Figure 3. A schematic of black hole and QED photosphere. For black holes with temperature $T_{\rm BH} >T_{\rm crit} \approx 45$GeV, the electrons, positrons, and photons emitted from the black hole travel a distance $\sim(\alpha^2T_{\rm BH})^{-1}$ before they begin to brems\-strahlung scatter. Eventually, the particles scatter enough that a near-perfect fluid photosphere forms at $r_{0}$ with some temperature $T_{0}$.  The temperature of the fluid decreases as it flows outward, and at $r_{p}$, $T=m_{0e}$ and most of  the $e^{\pm}$ annihilate. The photons then  free stream to infinity--now with an average energy of $\bar{E}_{\rm obs}\sim m_{0e}(T_{\rm BH}/T_{\rm crit})^{1/2}$.\\

\end{document}